\documentclass[%
 reprint,
nofootinbib,
 aps,
]{revtex4-1}
\usepackage{amsmath,amssymb}
\usepackage{graphicx}
\usepackage{color}
\usepackage{subfigure}

\newcommand{\Slash}[1]{{\ooalign{\hfil/\hfil\crcr$#1$}}} 

\makeatletter
\newcommand{\figcaption}[1]{\def\@captype{figure}\caption{#1}}
\newcommand{\tblcaption}[1]{\def\@captype{table}\caption{#1}}
\makeatother
\makeatletter

\@addtoreset{equation}{section}
\makeatother

\begin{document}
\baselineskip=16pt
\begin{titlepage}
\begin{flushright}
{\small OU-HET 724/2011}\\
\end{flushright}
\vspace*{1.2cm}

\begin{center}

{\Large\bf 
Parity violation in QCD process 
} 
\lineskip .75em
\vskip 1.5cm

\normalsize
{\large Naoyuki Haba}$^1$,
{\large Kunio Kaneta}$^1$,
{\large Shigeki Matsumoto}$^2$,\\
{\large Takehiro Nabeshima}$^3$, and 
{\large Soshi Tsuno}$^4$

\vspace{1cm}

$^1${\it Department of Physics, 
 Osaka University, Toyonaka, Osaka 560-0043, 
 Japan} \\

$^2${\it IPMU, TODIAS, University of Tokyo, Kashiwa, 277-8583, Japan}\\

$^3${\it Department of Physics, University of Toyama, Toyama 930-8555, Japan}\\

$^4${\it High Energy Accelerator Research Organization (KEK), 
Tsukuba, Ibaraki 305, Japan}

\vspace*{10mm}

{\bf Abstract}\\[5mm]
{\parbox{13cm}{\hspace{5mm}
%

Parity violation in QCD process is studied using 
 helicity dependent top quark pair
 productions at Large Hadron Collider experiment. 
Though no violation can be found in the standard model (SM), 
 new physics beyond the SM predicts the violation in general. 
In order to evaluate the violation, we utilize an effective 
 operator analysis in a case that
 new particles predicted by the new physics
 are too heavy to be directly detected. 
By using this method, we try to
 discriminate supersymmetric SM from
 universal extra-dimension model via an 
 asymmetry measurement
 of the top quark pair production. 
We also discuss the asymmetry from the SM electroweak top pair production process and that from the little Higgs model.
\newline

PACS numbers: 12.38.Bx,  12.60.-i, 12.60.Jv, 14.65.Ha
}}

\end{center}

\end{titlepage}

\section{Introduction}

The standard model (SM) successfully discribes  
 almost all experiments below ${\cal O}(100)$ GeV. 
However, 
 there are some difficulties in the SM,  
 for example, 
 a hierarchy problem in the Higgs sector 
 as a quantum field theory. 
It is also a problem that 
 the SM does not contain  
 natural candidate of dark matter.  
People believe an existence of 
 underlying theory beyond the SM 
 above a scale of ${\cal O}(1)$ TeV, 
 and its discovery is strongly expected 
 at Large Hadron Collider (LHC) experiment. 
A lot of candidates beyond the SM have  
 been suggested until now. 
The first promising candidate is 
 supersymmetry (SUSY), in which 
 the gauge hierarchy problem is naturally solved as well as 
 gauge coupling unification is achieved. 
In the SUSY with $R$-parity, 
 the lightest SUSY particle 
 is stable which can be a dark matter. 
The second candidate is 
 extra-dimension theory, 
 which might be related to 
 string theory.   
Solutions of the hierarchy problem 
 have been suggested in various 
 setup in the extra-dimension (see, for example, Ref.\cite{Randall:1999ee}).  
Here we consider an universal extra-dimension (UED) model as 
 the simplest example of 
 the extra-dimension theory.  
The UED model is 
 a naive extension of the SM\cite{Appelquist:2000nn}, 
 where SM particles have extra-dimensional modes, i.e., 
 Kaluza-Klein (KK) particles,
 and their spins are the same as SM particles.
The lightest KK particle is stable 
 and can be a dark matter. 
Both  SUSY and UED predict 
 new heavy particles
 as superpartners  and KK-particles, respectively.

For those models,
 it is important to experimentally distinguish one from 
 the other, not only by the mass spectrum but also by the kinematic feature
 in the production. 
Actually, the production processes between SUSY and UED
 models are similar at hadron colliders, so that the study of the kinematic 
 properly will become more important role to determine it. 
The spin correlation 
 will be the direct probe to determine the model, although we meet a difficulty 
 how to measure the spin-correlation in production\cite{Bernreuther:2008md}.
The aim of this paper is to discriminate SUSY from UED
 at the LHC by focusing on the 
 parity violation in QCD 
 processes without discovering any new particles.
 \footnote{The discrimination of SUSY and UED at the LHC is also 
 studied in \cite{Datta:2005zs}.}
We investigate the determination 
 of underlying theory through the 
 parity violation in QCD, 
 even if new particles are too heavy to be detected 
 directly in the experiments. 
Of cause the SM has no 
 parity violation in QCD, and 
 it can be happened 
 through the effects of 
 new physics.
In SUSY, 
 gluino-quark$_{L(R)}$-squark$_{L(R)}$ 
 interactions can violate parity,  
 since a mass of left-handed squark $\tilde{q}_{L}$ is different from 
 that of right-handed squark $\tilde{q}_{R}$ in general. 
While the UED has no parity violating QCD interactions 
 (at least in tree level). 
Therefore, we could distinguish SUSY from UED 
 through the QCD-parity violation. 
An asymmetry measurement of the helicity dependent 
 cross section of $t\bar{t}$ will suggests 
 the violation. 
The SM 
 effect will be also estimated, where  
 the asymmetry is caused by
 electroweak interactions
  in a tree-level
  {
  \cite{Beenakker:1993yr,Kao:1999kj,Moretti:2006nf,Bernreuther:2008md} . }
This is the background of the discovery of 
 parity violation in the QCD. 
We will also consider a little Higgs (LH) model as  
 the third candidate 
 beyond the SM, 
 where no parity violation exists in QCD but weak interactions 
 are modified from the SM. 
This will cause the asymmetry in 
 the helicity dependent cross sections of
 the 
 top pair production.

In this paper, 
 we study parity violation in QCD process by using 
 helicity dependent top
 quark pair productions at the LHC experiment. 
Though no violation can be found in the SM, 
 new physics beyond the SM predicts the violation in general. 
In order to evaluate the violation, we utilize an effective 
 operator analysis in a case that
 new particles predicted by the new physics
 are too heavy to be directly detected. 
By using this method, we try to
 discriminate SUSY SM from
 UED model via an 
 asymmetry measurement of the top quark pair production. 
We also discuss the asymmetry from the SM 
 electroweak top pair production process and that from the
 LH model.

\section{Top pair production with helicity dependence}

We investigate helicity dependent 
 top pair production at LHC experiment.
The helicity of top pair can be measured, 
 since  
 top quark immediately decays 
 before hadronization differently from 
 other quarks. 
The observed property in the decay products assessed the helicity 
information of the top quarks 
{
\cite{Stelzer:1995gc,Beneke:2000hk,Bernreuther:2001rq,Bernreuther:2004jv,Bernreuther:2006vg}
}.
{
Then a measurement of  
the cross section 
depending on helicities of top and anti-top
is possible. 
Here, at first, 
we show how to 
determine 
the helicity of top quark 
through 
an angular distribution of
charged lepton and momentum of multi-jets experimentally. 
As for a spin direction of a top quark, 
we focus on a case that
top quark decays into $l^+\nu b$ and anti-top is fully hadronic decay 
with jets. 
The top quark decays into $W^+b \to l^+\nu b$, 
where a helicity of $b$ is identified as 
a chirality of $b$ (i.e., $b_L$) since 
$m_b$ is negligible in this energy. 
In a rest frame of the top,  
this $W^+$ almost goes along the top spin axis 
and its polarization is longitudinal or $-1$, 
since the weak interaction is chiral. 
In this frame, longitudinal $W^+$ goes to $+1/2$ spin direction of the top, 
while left-handed $b$ goes to
the opposite direction. 
On the other hand, $-1$ polarized $W^+$ almost goes to 
$-1/2$ spin direction of the top. 
In both cases, $l^+$ almost emits in the $W^+$ polarization direction
due to angular momentum conservation.
Next step, we consider how to know the
direction of motion of the top. 
It can be measured since 
the anti-top quark decays into 3-jets which 
can be reconstructed. 
Therefore, we can determine the helicity of the top quark, and  
the helicity of anti-top can be also measured 
in a similar way.
}

If parity is violated in QCD through 
 the effects of new physics, 
 there must appear asymmetry in the cross section 
 depending on helicities of top and anti-top.  
For the asymmetry with the helicity dependence in the $t \bar{t}$ 
 mass system,
 $m_{t\bar t}$ distribution is defined as
\begin{align}
\delta A_{LR}^0(m_{t\bar{t}})
& = \frac{
\left(\frac{d\sigma_{++}}{d m_{t\bar t}}+
\frac{d\sigma_{+-}}{d m_{t\bar t}}\right)
-\left(\frac{d\sigma_{--}}{d m_{t\bar t}}
+\frac{d\sigma_{-+}}{d m_{t\bar t}}\right)}
{
\left(\frac{d\sigma_{++}}{d m_{t\bar t}}+
\frac{d\sigma_{+-}}{d m_{t\bar t}}\right)
+\left(\frac{d\sigma_{--}}{d m_{t\bar t}}
+\frac{d\sigma_{-+}}{d m_{t\bar t}}\right)}
 \; ,
\label{1}
\end{align}
where $\sigma_{\lambda_t\lambda_{\bar{t}}}$ denotes the production cross section with a top (anti-top) helicity $\lambda_t$ ($\lambda_{\bar t}$). 
We apply an effective operator analysis to the 
 top pair production up to 
 orders of $\alpha_W$ and $\alpha_s^2$. 
 {
We neglect a chirality flip via a top Yukawa coupling ($y_t$)
 that suggests an order of 
 $\alpha_s^2y_t$ in 1-loop diagrams.
 The analysis of this paper is the first step, and we will regard the effects of top Yukawa coupling in the next work\cite{Yukawa}.}
Then 
 $\sigma_{+-}$ and $\sigma_{-+}$ 
 in $\delta A_{LR}^0$ of Eq.(\ref{1}) 
 are irrelevant, 
 since 
 $\sigma_{+-}$ and $\sigma_{-+}$ are caused by
 chiral flip of top or anti-top quark. 
Namely, 
 the helicity of anti-top quark is automatically determined 
 by that of top quark 
 up to the order of $\alpha_s^2$.
For example, when 
 top quark helicity is $+$, helicity of the anti-top is
 also $+$ due to the no chirality flip. 
Then 
 $\sigma_{+-}$ and $\sigma_{-+}$ are redundant, 
 and it is useful to define 
 $\delta A_{LR}$ as 
\begin{equation}
\delta A_{LR}(m_{t\bar{t}})
 = \frac{\displaystyle{\frac{d\sigma_{++}}{d m_{t\bar t}}-\frac{d\sigma_{--}}{d m_{t\bar t}}}}
{\displaystyle{\frac{d\sigma_{++}}{d m_{t\bar t}}+\frac{d\sigma_{--}}{d m_{t\bar t}}}} \; . 
\label{1-1}
\end{equation}
We do not have to measure the helicity of anti-top 
 for $\delta A_{LR}$, and   
 we estimate $\delta A_{LR}$ of parton level
 in the following 
 analyses of investigating  
 the parity violation in QCD.

As for 
 a numerical calculation of quark and anti-quark
 annihilation and gluon fusion
 processes, we use GR@PPA event generator\cite{Tsuno:2006cu},
 where in calculation, we use the CTEQ6L1. 
The cross section is given by 
\begin{widetext}
\begin{align}
\sigma(pp\to t\bar{t})
=&\sum_{a,b}
\int ^1_0dx_1\int^1_0dx_2\hat{\sigma}(ab\to t\bar{t};\hat{s},x_1,x_2,\mu_R)
 D_{a/p}(x_1,M_Z)D_{b/p}(x_2,M_Z),
\end{align}
\end{widetext}
where 
 $D_{a/p}(x,\mu_F)$ is 
 a parton distribution function 
 with a  
 factorization scale of $\mu_F$, 
 which is chosen for $Z$ boson mass, 
 for simplicity. 
The $a$ and $b$ stand for gluon and quark flavor in the proton.  
The $\hat{\sigma}$ is a parton level cross section
 with an invariant mass of $a$ and $b$
 as $\hat{s}=(p_a+p_b)^2$ and scaling parameter $x$.
The $\mu_R$ is a renormalization scale which we take $M_Z$.

\section{SUSY and UED discrimination via QCD parity violation}

At first, 
 we represent 
 parity violating dimension 6 operators, 
 and, next, 
 we investigate the QCD parity violation 
 in SUSY and UED. 
Let us try to discriminate 
 SUSY from UED through 
 the parity violation even when 
 masses of sparticles or KK-particles 
 are too heavy to be detected at 
 direct searches. 
We also estimate 
 effects of weak parity violation 
 in the 
 SM and LH. 

\subsection{Dimension 6 operators}

We use an effective theory 
 where particles of new physics, such as, sparticles and KK-particles,
 are integrated out.   
The QCD parity violation is represented 
 by the SM field contents with 
 dimension 6 operators as a leading order. 
These irrelevant 
 operators in QCD are
 shown by 
 ${\cal
 O}^{(1)}_{qqqq}, {\cal O}^{(8)}_{qqqq}, {\cal O}_{qqG}$, and 
 ${\cal O}_{qqGG}$, 
 which represent color-singlet 4 fermi, color-octet 4 fermi,
 quark-quark-gluon, and quark-quark-gluon-gluon operators,
 respectively.
They are listed in Ref\cite{Haba:2011vi}, and given by 
\begin{widetext}
\begin{align}
{\cal O}^{(1)}_{qqqq}
=&\frac{12g_s^4}{192\pi^2}
\sum_{i,j=L,R}
\int
\frac{d^4k_1d^4k_2d^4k_3d^4k_4}{(2\pi)^4(2\pi)^4(2\pi)^4(2\pi)^4}
(2\pi)^4\delta^4(-k_1+k_2-k_3+k_4)\nonumber\\
&\qquad \qquad \qquad\times
C_{ij}
\left[
\bar{q}(k_1)\gamma^\mu P_iq(k_2)
\right]
\left[
\bar{q}'(k_3)\gamma_\mu P_jq'(k_4)
\right],\label{6}\\
{\cal O}^{(8)}_{qqqq}
=&\frac{12g_s^4}{192\pi^2}
\sum_{i,j=L,R}
\int
\frac{d^4k_1}{(2\pi)^4}
\frac{d^4k_2}{(2\pi)^4}
\frac{d^4k_3}{(2\pi)^4}
\frac{d^4k_4}{(2\pi)^4}
(2\pi)^4\delta^4(-k_1+k_2-k_3+k_4)\nonumber\\
&\qquad \qquad\qquad \times
D_{ij}
\left[
\bar{q}(k_1)T^a\gamma^\mu P_iq(k_2)
\right]
\left[
\bar{q}'(k_3)T^a\gamma_\mu P_jq'(k_4)
\right],\label{6-2}
\\
{\cal O}_{qqG}
=&\frac{g_s^3}{96\pi^2}
\sum_{i=L,R}
\int\frac{d^4k_1}{(2\pi)^4}\frac{d^4k_2}{(2\pi)^4}\frac{d^4k_3}{(2\pi)^4}
(2\pi)^4\delta^4(-k_1+k_2+k_3)
\bar{q}(k_1)T^aE^\mu_i G^a_\mu(k_3)P_iq(k_2),
\\
{\cal O}_{qqGG}
=&\frac{g_s^4}{192\pi^2}
\sum_{i=L,R}
\int\frac{d^4k_1}{(2\pi)^4}\frac{d^4k_2}{(2\pi)^4}\frac{d^4k_3}{(2\pi)^4}\frac{d^4k_4}{(2\pi)^4}
(2\pi)^4\delta^4(-k_1+k_2+k_3+k_4)\nonumber\\
&\qquad \qquad \qquad\times
\bar{q}(k_1)
\left[F^{\mu\nu}_i\delta^{ab}+H^{\mu\nu}_iT^aT^b\right]
G^a_\mu(k_2)G^b_\nu(k_3)P_iq(k_4),\label{7}
\end{align}
\end{widetext}
where $P_i\;{ (i=L,\;R)}$ is the chirality projection, $P_{L}=\frac{1-\gamma^5}{2} \left(P_{R}=\frac{1+\gamma^5}{2} \right)$, and
$E^\mu_i,\; F^{\mu\nu}_i,\; H^{\mu\nu}_i$ are defined as
\begin{align}
E^\mu_i
&
=\{e_{1i}\Slash{k}_1+e_{2i}\Slash{k}_2\}k^\mu_1
+\{e_{3i}\Slash{k}_2+e_{4i}\Slash{k}_1\}k^\mu_2\nonumber\\
&+\{e_{5i}k_1^2+e_{6i}k_2^2-e_{7i}k_1\cdot k_2\}\gamma^\mu\nonumber\\
&-e_{8i}i\epsilon^{\alpha\beta\mu\nu}\gamma_5\gamma_\nu
k_{1\alpha}k_{2\beta},\label{4}\\
F^{\mu\nu}_i&=
f_{1i\alpha} i\epsilon^{\alpha\mu\nu\beta}\gamma_5\gamma_\beta
+f_{2i\alpha}g^{\mu\nu}\gamma^\alpha
+f_{3i\alpha}g^{\alpha\mu}\gamma^\nu\nonumber\\
&+f_{4i\alpha}g^{\alpha\nu}\gamma^\mu,\label{2}
\end{align}
\begin{align}
H^{\mu\nu}_i&=
h_{1i\alpha} i\epsilon^{\alpha\mu\nu\beta}\gamma_5\gamma_\beta
+h_{2i\alpha}g^{\mu\nu}\gamma^\alpha
+h_{3i\alpha}g^{\alpha\mu}\gamma^\nu\nonumber\\
&+h_{4i\alpha}g^{\alpha\nu}\gamma^\mu.\label{3}
\end{align}
In Eqs.(\ref{6-2}) and (\ref{4}),
 ${ C_{ij}, D_{ij},} e_{1i}, \cdots {, e_{8i}}$
 are Wilsonian coefficients which have mass dimension
 $M^{-2}_{\rm NP}$, where 
 $M_{\rm NP}$ stands for
 a scale of new physics characterized by 
 new particles' masses. 
In Eqs.(\ref{2}) and (\ref{3}), 
 coefficients $f_{1i\alpha}, \cdots, h_{1i\alpha}, \cdots$
 are some combination of the quark and the gluon momentum,
 e.g.
\begin{align}
 f_{1L\alpha}
&
=f^{(k_1)}_{1L}k_{1\alpha}+f^{(k_3)}_{1L}k_{3\alpha}+f^{(k_4)}_{1L}k_{4\alpha}
\end{align}
for left-handed quarks, and $f^{(k_1)}_{1i}, \cdots, f^{(k_4)}_{4i}$, 
$h^{(k_1)}_{1i}, \cdots, h^{(k_4)}_{4i}$ are Wilsonian coefficients with
mass dimension $M^{-2}_{\rm NP}$. 
The QCD interactions in Eqs.(\ref{6})-(\ref{7}) become chiral 
 in the SUSY SM, since their coefficients are different between
 left- and right-handed quarks.
We can see explicit coefficients 
 of these operators in the SUSY SM and UED model 
 (and also LH model) in Ref.\cite{Haba:2011vi}.

\subsection{SUSY}

In the SUSY SM with $R$-parity, 
 SUSY particles can propagate only inside loop diagrams, and 
 the maximal contributions of the parity violation come from 
 1-loop induced dimension 6 operators.  
There were some estimations previously, 
 where SUSY particles have masses of
 ${\mathcal O}(100)$ GeV{
 \cite{Hollik:1997fd,Berge:2007dz,Sullivan:1996ry}.
 } 
Especially, in Ref.\cite{Berge:2007dz},
 the asymmetry was estimated as 
 $|\delta A_{LR}(m_{t\bar{t}})|\simeq 2.0$\% 
 with ${\mathcal O}(100)$ GeV sparticles. 
Here we show a similar estimation by use of 
 dimension six operators 
 by integrating out heavy SUSY 
 particles. 
We neglect 
 the left-right mixing 
 in the stop mass matrix, which   
 corresponds to neglect top Yukawa 
 in the loop level. 

We should take mass bounds of 
 gluino and squarks 
 constrained by LHC experiment\cite{daCosta:2011qk}.
Cross sections from SUSY dimension 6 operators 
at a center-of-mass energy $E_{\rm CM}=7$ TeV
 are listed in Table \ref{Xs_SUSY}. 
Where we take some sample points of sparticle masses as 
 $(m_{\tilde{g}}, m_{\tilde{t}_L}, m_{\tilde{t}_R})$
$ = (2000, 2100, 1000),$ $(2000, 1200, 1000),$ and $(400, 1200, 410)$ GeV.
Cases of (i) and (ii) show heavy sparticles consistent with 
 LHC data\cite{daCosta:2011qk}.
In case of (iii), parameter set shows
 gluino and one of stop are degenerate within 30  GeV,
 which is not excluded experimentally, too. 
It is because 
 there are experimental cut for $p_T$s of multi-jets 
 with missing transverse momentum 
 in SUSY search at LHC (Tevatron), where
 an event selection for jets is
 $p_T>40$ GeV\cite{daCosta:2011qk} ($p_T>30$ GeV\cite{:2007ww}), 
 and 
 $p_T$ of jets are roughly estimated as the mass difference
 of gluino and squarks. 
For calculating a cross section,
 we should pay attention to the cut of collision energy at parton level.
In the effective operator approach,
 operators are expanded by sparticle masses.
This means the parton level invariant mass $\sqrt{\hat{s}}$ can not be
 larger than
 sparticle mass, $\sqrt{\hat{s}}<M_{\rm SUSY}$.
Here, we estimate the cross section under the limit of
 $\sqrt{\hat{s}}\leq 0.95 \times M_{\rm SUSY}$, where 
 $M_{\rm SUSY}$ stands for Min.$[m_{\tilde{g}},
 m_{\tilde{t}_L}, m_{\tilde{t}_R}]$.
Note that 
 top and anti-top are mainly produced in QCD
 process in collider experiment 
 and the experimental data shows 
 $\sigma^{\rm exp}(pp\to t\bar t)=179$ pb\cite{totalcross}, 
 where 
 the magnitude of $\sigma^{\rm exp}$ is 
 almost same as that induced from the SM QCD processes,  
 $\sigma^{\rm SM}$. 
Table \ref{Xs_SUSY} shows 
 cross sections of (i) and (ii) 
 are (roughly) $10^{-6}$ smaller than $\sigma^{\rm SM}$. 
Even if gluino mass is ${\cal O}(100)$ GeV as the case of (iii),
 the cross section does not drastically increase. 
We show the numerical calculation of
 the asymmetry $\delta A_{LR}$ for the case of
 (i) 
together with UED result in subsection 3.5. 

\begin{figure}[h]
  \def\@captype{table}
  \begin{minipage}[c]{.48\textwidth}
   \begin{center}
    \begin{footnotesize}
     \begin{tabular}[t]{|c|c|c|}
      \hline
 $(m_{\tilde{g}}, m_{\tilde{t}_L}, m_{\tilde{t}_R})$ [GeV]&
 $\sigma^{\rm SUSY}$ [$10^{-4}$pb]
 &$\Delta \sigma$ [$10^{-2}$pb] \\\hline\hline
 (2000, 1010, 1000)& 2.5&-5.2\\\hline
 (2000, 1100, 1000)& 2.3&-5.0\\\hline
 (2000, 1200, 1000)& 2.2&-4.9\\\hline
 (2000, 1300, 1000)& 2.3&-4.7\\\hline
 (2000, 1400, 1000)& 2.0&-4.6\\\hline
 (2000, 1500, 1000)& 1.9&-4.5\\\hline
 (2000, 1600, 1000)& 1.8&-4.3\\\hline
 (2000, 1700, 1000)& 1.8&-4.1\\\hline
 (2000, 1800, 1000)& 1.7&-4.0\\\hline
 (2000, 1900, 1000)& 1.6&-3.8\\\hline
 (2000, 2100, 1000)& 1.5&-3.6\\\hline
 (2000, 2200, 1000)& 1.5&-3.6\\\hline
 (2000, 2300, 1000)& 1.5&-3.5\\\hline
     \end{tabular}
\tblcaption{Sparticles masses and corresponding cross
 sections are listed. 
 $\sigma^{\rm SUSY}\equiv\sigma^{\rm SUSY}(pp\rightarrow t\bar{t})$
 and $\Delta \sigma\equiv \sigma^{\rm SM + SUSY}-\sigma^{\rm SM}$.
}
     \label{xs_mdiff}
    \end{footnotesize}
    \end{center}
  \end{minipage}
  \hfill
  \begin{minipage}[c]{.48\textwidth}
   \includegraphics[width=200pt]{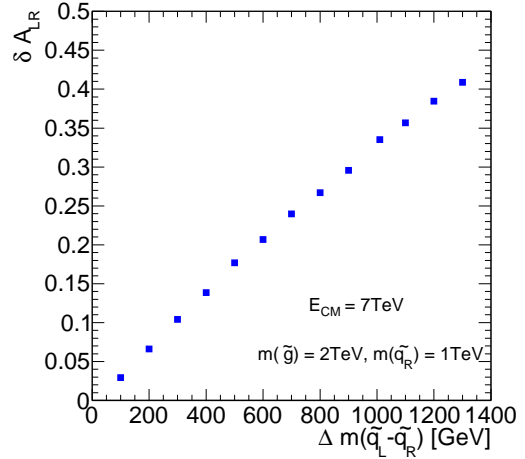}
\caption{The relation between $\delta A_{LR}$ and the mass differences 
 in Table 2. 
The horizontal axis denotes   
 $\Delta m(\tilde{q}_L-\tilde{q}_R)\equiv m_{\tilde{q}_L}-m_{\tilde{q}_R}$.}
\label{susyasym_mdiff}
  \end{minipage}
\end{figure}

{
Table \ref{xs_mdiff} shows 
cross sections at $E_{\rm CM}=7$ TeV
with various magnitudes of $m_{\tilde{t}_L}$ fixing $m_{\tilde{g}}$ and $m_{\tilde{t}_R}$.
In Table \ref{xs_mdiff}, $\Delta \sigma$ is defined as $\Delta \sigma\equiv \sigma^{\rm SM+SUSY}-\sigma^{\rm SM}$,
where $\sigma^{\rm SM+SUSY}$ stands for a cross section including QCD, SM electroweak (SMEW), and SUSY contributions.
The SUSY contribution is small, i.e., cross section is ${\cal O}(10^{-4})$ pb and $| \Delta \sigma|\simeq {\cal O}(10^{-2})$ pb.}
The values of $\delta A_{LR}$
 depending on the masses 
 are shown in 
 Fig. \ref{susyasym_mdiff}.  
Apparently, the larger the
 mass difference between $\tilde{q}_L$ and 
 $\tilde{q}_R$ becomes, 
 the larger magnitude of $\delta A_{LR}$ becomes. 
%

 \begin{table}[htbp]
\begin{center}
\begin{tabular}{|l|c|}
\hline
 $(m_{\tilde{g}}, m_{\tilde{t}_L}, m_{\tilde{t}_R})$ [GeV]&$\sigma^{\rm SUSY}(pp\rightarrow t\bar{t})$ [$\times 10^{-4} $pb]\\\hline\hline
 (i), (2000, 2100, 1000)&1.5\\\hline
 (ii), (2000, 1200, 1000)&2.2\\\hline
 (iii), (400, 1200, 410)&8.8\\\hline
\end{tabular}
\caption{\label{Xs_SUSY} Sample points of sparticle masses and their  cross sections at $E_{\rm CM}=7$ TeV}
\end{center}
\end{table}

\subsection{UED}

In the UED with KK-parity, KK particles can propagate only inside loop
 diagrams, 
 and the maximal contributions come from 1-loop induced dimension 6 operators.
The QCD parity violation is induced 
 only through 
 non-degeneracy of KK masses 
 induced from radiative corrections. 
The dimension six operators 
 are listed in Ref.\cite{Haba:2011vi}, which 
 contribute
 both $q\bar q$ annihilation processes and 
 gluon fusion subprocesses. 
In UED we take the cut of collision
 energy to $\sqrt{\hat{s}}\leq 1/R$, where $R$ is the compactification
 scale. 
At tree level, KK particles are degenerate, 
 but there appears a slight difference between the left-handed KK quark mass,
 $m^{(n)}_L$, and right-handed one, $m^{(n)}_R$, through the
 renormalization effects \cite{Cheng:2002iz} as    
\begin{align}
\delta m_{t^{(n)}_L}
&=\left(\frac{n}{R}\right)
\left(
3\frac{g_s^2}{16\pi^2}+\frac{27}{16}\frac{g^2}{16\pi^2}+\frac{1}{16}\frac{g'^2}{16\pi^2}
\right)\log \frac{\Lambda^2}{\mu^2},\\
\delta m_{t^{(n)}_R}
&=\left(\frac{n}{R}\right)
\left(
3\frac{g_s^2}{16\pi^2}+\frac{g'^2}{16\pi^2}
\right)\log \frac{\Lambda^2}{\mu^2},
\end{align}
where $\Lambda$ and $\mu$ are the cutoff scale and the renormalization
 scale, respectively.
Accurately speaking, 
 these effects are beyond the order of $\alpha_s^2$, 
 however, here we take them into account, since 
 it might be informative. 
We take $\mu=\sqrt{\hat{s}}\geq 2m_t$, and helicity asymmetry is plotted
 in a sample point as 
 $(R^{-1}, \Lambda)=(2\ {\rm TeV},\; 20\ {\rm TeV})$.
Fixing the $\Lambda R$ means that KK-mode appears up to $n=20$
 below the cutoff scale.
Here we take the sum of KK-modes up to infinity, for simplicity,  
 because a difference of coefficients between
 the sum of $n$ up to 20 and infinity is less than
 3\%.   
A numerical result of the
 magnitude of $\delta A_{LR}$ in the UED model  
 will be shown 
 in Fig.\ref{mtt_asym} and \ref{pt_asym} in subsection 3.5.

\subsection{SM electroweak background and LH}

Here we estimate $\delta A_{LR}$ induced from 
 not QCD but weak interactions, which 
 is the   
SMEW background. 
The  SMEW background is not negligible, 
 although it was not estimated in Ref.\cite{Berge:2007dz}. 
The asymmetry from the SMEW, $\delta A_{LR}^{\rm SMEW}$,  
 is induced by 
 electroweak interactions
 at a tree-level, and  
 the cross section of the SMEW 
 is estimated as 
 $\sigma^{\rm SMEW}(pp\rightarrow t\bar{t}) \simeq 3.4\times 10^{-1} {\rm pb}$.
 \footnote{
Input parameters are  
$(g_s, g, g')=(1.3, 0.65, 0.31),\;M_Z=91\;{\rm GeV},$
 and
 $m_t=173\;{\rm GeV}$.
} 
{
The SMEW contribution is larger than the SUSY contribution comparing to Table \ref{Xs_SUSY}.
It is worth noting that a magnitude of $\sigma^{\rm QCD+SMEW}(\simeq$125 pb) is smaller than that of only QCD contribution 
$\sigma^{\rm QCD}(\simeq138$ pb).
The SMEW contributions in $pp\to t\bar t$ process is studied in Ref.\cite{Kuhn:2005it,Kuhn:2006vh,Moretti:2006ea}.
}

As for the LH, 
 there is no QCD parity violation as 
 in the SM. 
The effects of the LH 
 is summarized as deviated 
 weak interactions from the SM. 
Integrating out 
 new heavy particles in the LH, 
 the effective 
 4 Fermi operators are 
 induced as \cite{Haba:2011vi} 
\begin{align}
&{\cal O}^{\rm LH}
=\frac{g^2}{2}\cos^2\beta\frac{1}{k^2_W-M_W^2}
(\bar{t}\gamma^\mu P_Lb)(\bar{b}\gamma^\mu P_Lt)\nonumber\\
&+\sum_{q}^{\rm flavor}\frac{g^2}{3}\tan^2\theta_W\cos^2\beta\frac{1}{k^2_Z-M_Z^2}
(\bar{q}\gamma^\mu P_Lq)(\bar{t}\gamma^\mu P_Lt),
\end{align}
where $\cos \beta\sim 1-\frac{v^2}{2f^2}$,
 $g$ is the $SU(2)_L$ coupling, and 
 $v$ ($f$) is a vacuum expectation value (VEV) of the Higgs
 in the SM (LH), respectively. 
$k_W$ and $k_Z$ stand for 
 momenta of $W$ and $Z$ bosons, 
 respectively. 
When the LH takes VEV as 
 $f= 2$ TeV, 
 the angle $\beta$ becomes
 $\cos \beta = 0.992$.  
We will estimate the helicity asymmetry in the LH
 in Figs. \ref{mtt_asym} and \ref{pt_asym}. 
Note that the cross section of the LH model is the same order of 
 that of the SMEW processes 
 as $\sigma^{\rm LH}(pp\to t\bar t)=2.4\times 10^{-1}$ pb.

\subsection{Discriminate SUSY from UED}

\begin{figure}[h]
  \def\@captype{table}
  \begin{minipage}[c]{.48\textwidth}
  \includegraphics[width=80mm]{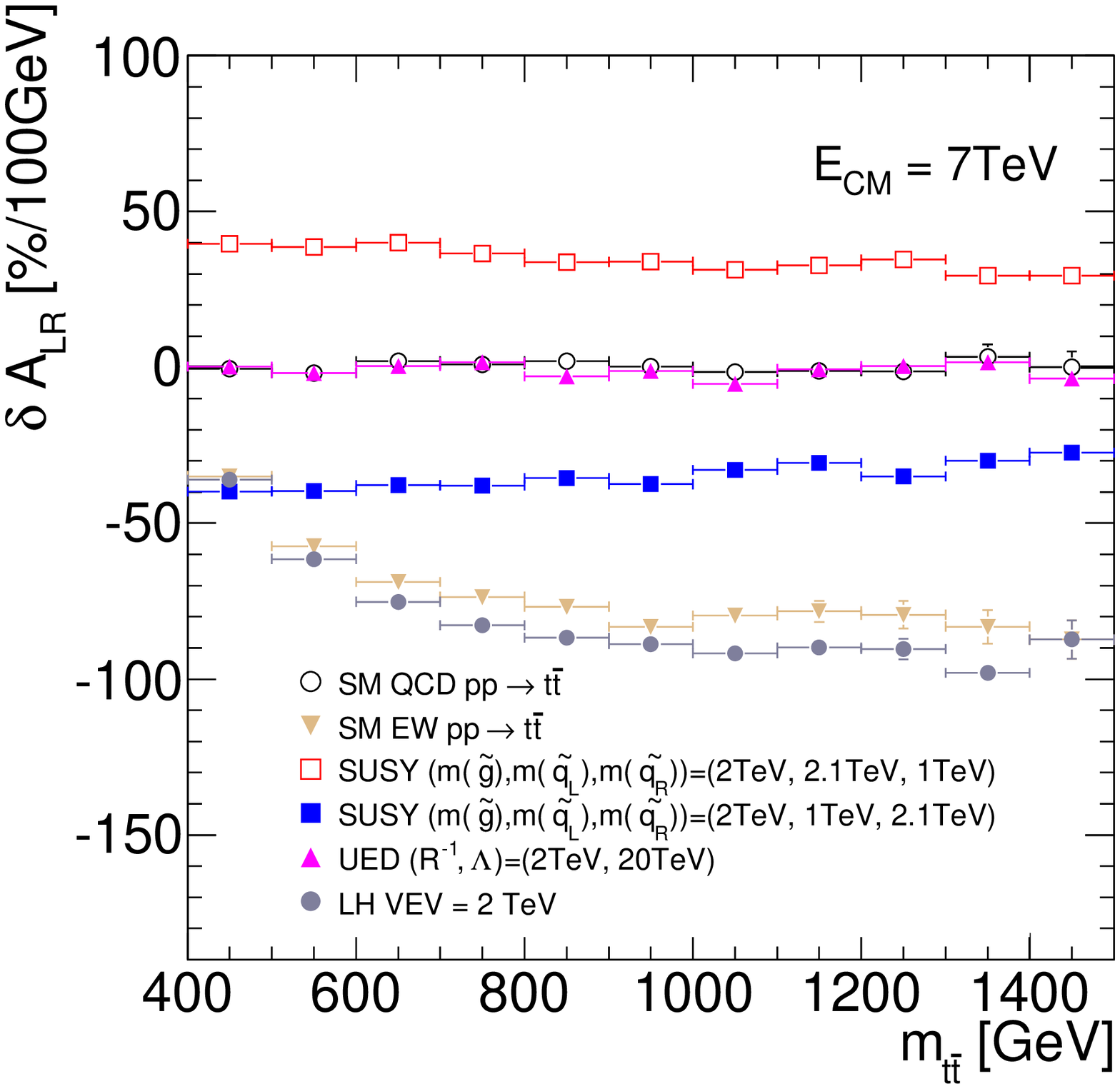}
 \caption{\footnotesize $\delta A_{LR}$ of $m_{t\bar t}$ distribution.} 
 \label{mtt_asym}
\end{minipage}
  \hfill
  \begin{minipage}[c]{.48\textwidth}
  \includegraphics[width=80mm]{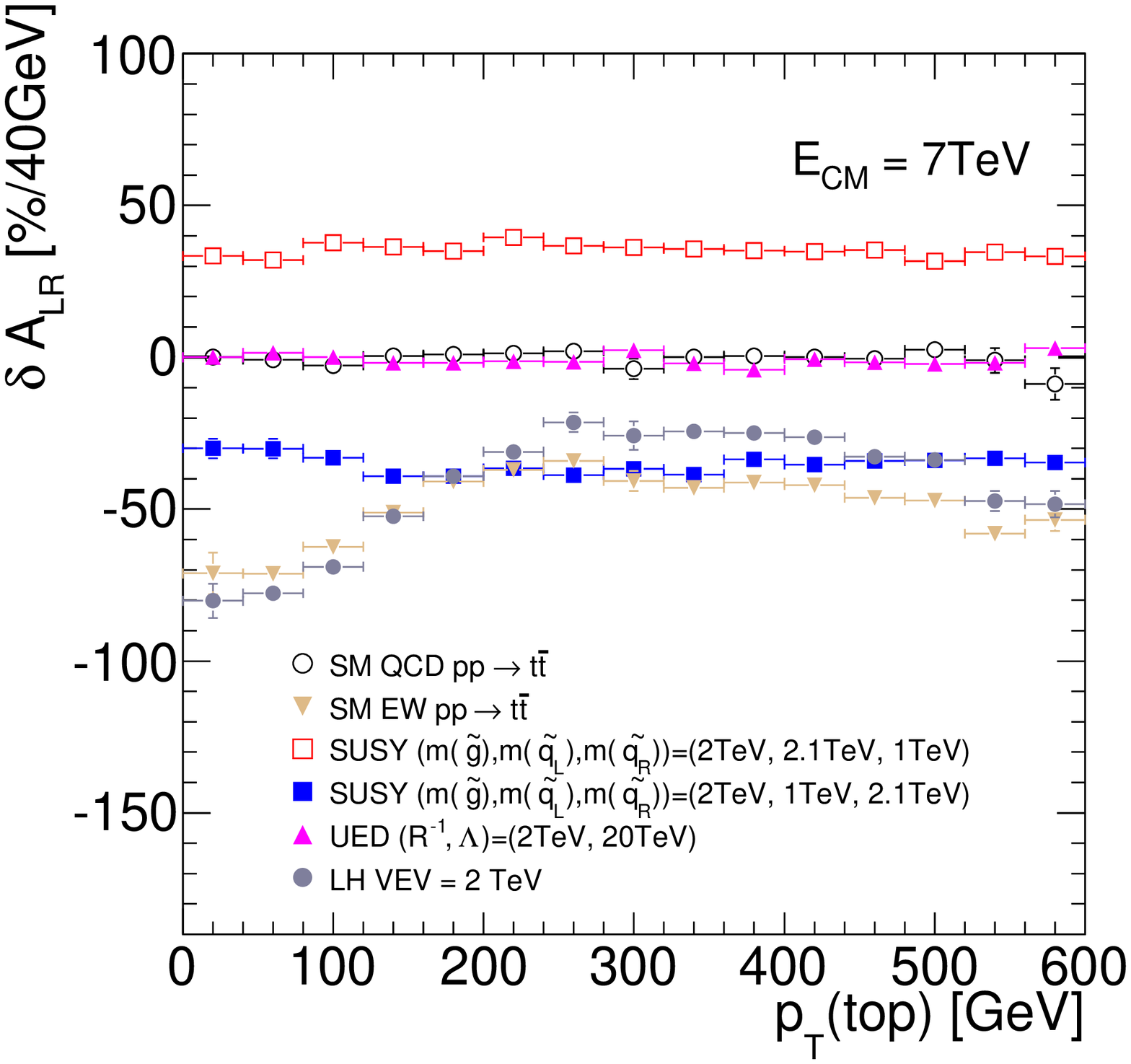}
  \caption{\footnotesize Dependence of $\delta A_{LR}$ on $p_T$.}
  \label{pt_asym}
\end{minipage}
 \end{figure}

\begin{figure}[h]
 \def\@captype{table}
  \begin{minipage}[c]{.48\textwidth}
  \includegraphics[width=80mm]{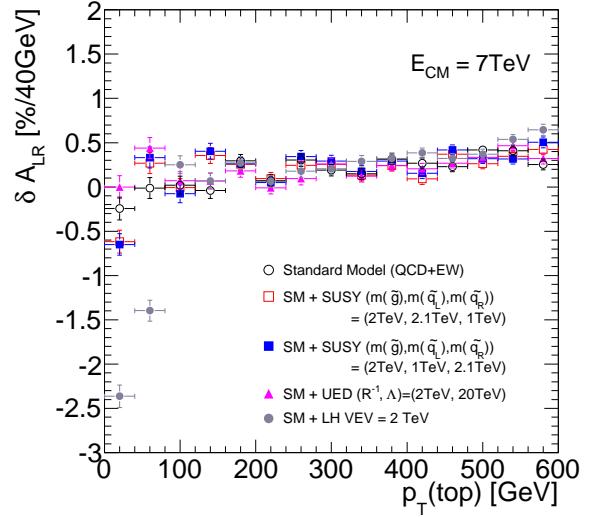}
  \caption{\footnotesize Dependence of $\delta A_{LR}$ on $p_T$ with SM interference.}
  \label{SM_NP_pT}
\end{minipage}
\end{figure}

Figures \ref{mtt_asym} $\sim$ \ref{SM_NP_pT}
 show results of numerical analyses of 
 the magnitude of $\delta A_{LR}$ 
 in the SUSY, UED, SMEW, and LH 
 depending on 
 $m_{t\bar{t}}$ and $p_T$, respectively. 
For example,  
 $\delta A_{LR}$ in SUSY,
 denoted by $\delta A_{LR}^{\rm SUSY}$, 
 is 
 defined by 
 $\delta A_{LR}^{\rm SUSY}
 = \frac{{d\sigma_{++}^{\rm SUSY}}/{d m_{t\bar t}}
 -{d\sigma_{--}^{\rm SUSY}}/{d m_{t\bar t}}}
 {{d\sigma_{++}^{\rm SUSY}}/{d m_{t\bar t}}
 +{d\sigma_{--}^{\rm SUSY}}/{d m_{t\bar t}}}$,   
 while 
 an observable magnitude of $\delta A_{LR}$ in 
 the experiments is given by 
 $\delta A_{LR}^{\rm exp} 
 = \frac{{d\sigma_{++}^{\rm exp}}/{d m_{t\bar t}}
 -{d\sigma_{--}^{\rm exp}}/{d m_{t\bar t}}}
 {{d\sigma_{++}^{\rm exp}}/{d m_{t\bar t}}
 +{d\sigma_{--}^{\rm exp}}/{d m_{t\bar t}}}$
{
, where $\sigma^{\rm exp}$ is the total cross section
.

 The magnitude of 
 $\delta A_{LR}^{\rm SUSY}$ in Figs. \ref{mtt_asym} and \ref{pt_asym} 
 does not include an interference of SM and SUSY.
 }
 $\delta A_{LR}$ of UED, SMEW, and LH in Figs. \ref{mtt_asym} and \ref{pt_asym} 
 are defined 
 in the same way. 
 We input SUSY mass parameters
 as $(m_{\tilde{g}}$, $m_{\tilde{q}_L}$, $m_{\tilde{q}_R})$ $=$ $(2{\rm TeV}$, $2.1{\rm TeV}$, $1{\rm TeV})$,
 and $(2{\rm TeV}$, $1{\rm TeV}$, $2.1{\rm TeV})$. 
Parameters of UED model are taken as
 $(R^{-1}, \Lambda)=(2{\rm TeV}, 20{\rm TeV})$. 
Apparently, 
 the helicity asymmetry in the SUSY SM can be larger than that
 in UED model, which  
 is of cause due to the squark mass splitting. 
For example, 
 when 
 $m_{\tilde{t}_L}\gg m_{\tilde{t}_R}$, 
 left-handed top pair production should be suppressed, 
 and then 
 the sign of 
$\delta A_{LR}^{\rm SUSY}$
 becomes positive 
 because $\sigma_{++}$ is larger than $\sigma_{--}$.
The opposite case is similarly understood. 
 \begin{figure}[h]
 \def\@captype{table}
  \begin{minipage}[c]{.48\textwidth}
  \includegraphics[width=80mm]{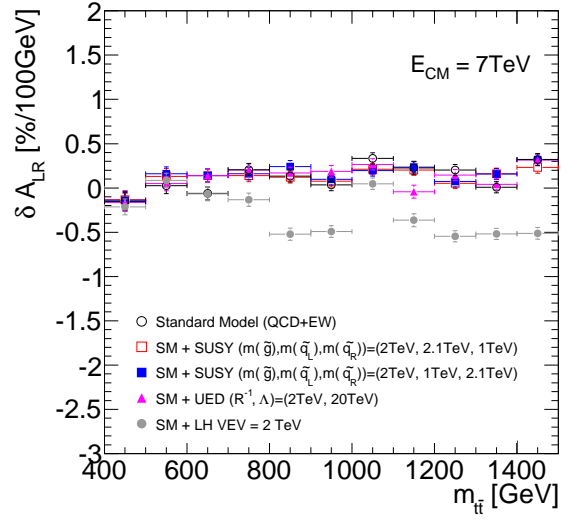}
  \caption{\footnotesize Dependence of $\delta A_{LR}$ on $m_{t\bar t}$ with SM interference.}
  \label{SM_NP}
\end{minipage}
\end{figure}
%
However, 
 we should notice that 
 the SUSY cross section is much smaller than that of the 
 SM QCD, and  
 unfortunately,  
 once the SM 
 interferes the 
 SUSY contribution, 
 the asymmetry could not seen by
 the large SM QCD contribution.
 
 {
Figures \ref{SM_NP} and \ref{SM_NP_pT} show $\delta A_{LR}$ including the interference of SM and the new physics (SUSY, UED and LH),
where $\delta A_{LR}^{\rm SM+SUSY}$ is around $3\times 10^{-3}$, 
and a deviation of $\delta A_{LR}^{\rm SM+SUSY}$ from $\delta A_{LR}^{SM}$ is roughly estimated as $1\times 10^{-3}$.
Here SM means QCD + SMEW.
As for LH, there is a large contribution as $\delta A_{LR}^{\rm LH}\sim 5 \times 10^{-3}$.
Since the helicity asymmetry is measured by 
 a spin correlation, $\delta A_{LR}^{\rm exp}$ should be larger 
 than an error of the spin correlation for observation. 
Thus, $\delta A_{LR}^{\rm SM + SUSY}(\sim 3\times 10^{-3})$ is difficult to be observed.
For example, in Ref.\cite{atral117}, 
 a correlation coefficient in a helicity basis 
 is represented, and the statistic and the systematic 
 errors are of order 0.1. 
However, we could expect that the statistic error reduces about 
 1/3 by 
 10 times events in the future LHC experiments
 and the systematic error 
 reduce about 1/10. 
 In the rest of this section, we discuss a potential to observe $\delta A_{LR}$.}
 
Note that 
 the cross section of the SM QCD decreases comparing 
 to that of the SUSY SM in high $m_{t\bar{t}}$ and $p_T$ region, 
 since  
 SUSY interactions are represented by irrelevant operators.   
So the SUSY signal could be significant, if 
 we select the phase space (final states) 
 as well as 
 take cut to focus on 
 high $m_{t\bar t}$ and $p_T$ region, 
 where 
 the SM QCD contribution 
 should be suppressed. 
 {
Tables \ref{qqtt} and \ref{ggtt} show 
 fractions of $(\lambda_t,\lambda_{\bar t})=(+,-)/(-,+)$, $(+,+)$, and $(-,-)$
 for SM QCD, SMEW, SUSY-L, and SUSY-R at $E_{\rm cm}=7$TeV. 
 Where SUSY-L and SUSY-R stand for $(m_{\tilde{g}},m_{\tilde{q}_L},m_{\tilde{q}_R})=(2 {\rm TeV},2.1 {\rm TeV},1 {\rm TeV})$ and $(2 {\rm TeV},1 {\rm TeV},2.1 {\rm TeV})$, respectively.
Notice that SM QCD contribution of $(+,+)$ and $(-,-)$ are suppressed in $gg\to t\bar t$ process.}
 \begin{table}[htbp]
\begin{center}
\begin{tabular}{|c|c|c|c|c|}
\hline
Helicities&SM QCD&SMEW&SUSY-L&SUSY-R\\\hline\hline
$(+, -)/(-, +)$&0.222&0.181&0.022&0.023\\\hline
$(+, +)$&0.385&0.178&0.570&0.402\\\hline
$(-, -)$&0.393&0.641&0.408&0.575\\\hline
\end{tabular}
\caption{\label{qqtt} Helicity fractions in $q\bar q \to t \bar t$ process}
\end{center}
\end{table}
 \begin{table}[htbp]
\begin{center}
\begin{tabular}{|c|c|c|c|c|}
\hline
Helicities&SM QCD&SMEW&SUSY-L&SUSY-R\\\hline\hline
$(+, -)/(-, +)$&0.747&---&0.000&0.000\\\hline
$(+, +)$&0.125&---&0.715&0.283\\\hline
$(-, -)$&0.127&---&0.285&0.717\\\hline
\end{tabular}
\caption{\label{ggtt} Helicity fractions in $gg \to t \bar t$ process}
\end{center}
\end{table}
Moreover, when top Yukawa coupling is large, 
 the parity
 violation could be enhanced 
 since 1-loop diagrams with 
 a charged Higgs(ino) inside the loop 
 has the top Yukawa couplings, 
 which have  
 $t_R$ and $\bar{t}_R$ in the external lines 
 (while 
 bottom Yukawa couplings 
 have $t_L$ and $\bar{t}_L$ in the external lines).
This effect is order of 
 $\alpha_s y_t^2$, which could be 
 the same order as $\alpha_s^2$ in 
 small $\tan\beta$ region. 
Then, 
 if the SUSY cross section is enhanced
 as $\sim 10^{-2}$ pb 
 at the high $m_{t\bar t}$ and $p_T$ region 
 with the specific phase space 
 (where 
 the SM QCD cross section 
 could be suppressed as $\sim 1$ pb), 
 $\delta A_{LR}^{\rm SUSY}$ could be large 
 enough 
 as $0.05$. 
 {
 In this case, $\delta A_{LR}$ can be observed when the statistic error is of oder 0.01.
 We need $10^{4}$ events for this statistic error.
 This number of events is the difference between $(+, +)$ and $(-, -)$ in $t\bar t$ production.
 Then, the total event should be $2\times 10^5$ to obtain $\delta A_{LR}\simeq0.05\pm 0.01$,
 and an integrated luminosity is roughly estimated as $10^2\;{\rm fb}^{-1}$.
 }
Therefore, 
 we should take reanalyses of $\delta A_{LR}$ 
 up to ${\cal O}(\alpha_s y_t^2)$ 
 with no use of dimension 6 operators,  
 and 
 we need 
 more detailed studies 
 for the discrimination between the SUSY SM and UED model.

As for the LH model, 
 $\delta A_{LR}^{\rm LH}$ 
 is the same order as $\delta A_{LR}^{\rm SMEW}$ 
 as shown in Figs. \ref{mtt_asym} and \ref{pt_asym}. 
Their asymmetries could be observable 
 when the SM QCD cross section is suppressed,  
 where the LH might be also discriminated from the SMEW 
 in a specific value of $f$.

\section{Summary and discussions}

In this paper, 
 we have studied parity violation in QCD process by using
 helicity dependent top
 quark pair productions at the LHC experiment. 
Though no violation can be found in the SM, 
 new physics beyond the SM predicts the violation in general. 
In order to evaluate the violation, we have utilized an effective 
 operator analysis in a case that 
 new particles predicted by the new physics
 are too heavy to be directly detected. 
By using this method, we have tried to
 discriminate SUSY SM model from
 UED model via an 
 asymmetry measurement of the top quark pair production. 
{
In spite of the tiny asymmetries of the SUSY and UED,}
 there are still possibilities of the discrimination to 
 succeed, i.e.,  
 we will take the analyses of order $\alpha_s y_t^2$ 
 in the small $\tan\beta$ region, and 
 investigate without use of the effective operators 
 in the specific phase space{\cite{Yukawa}}.  
We have also estimated the asymmetries from 
 the SMEW 
 background and the LH model.
They are the same order and 
 could be observable in the specific phase space.
{
For other models, which might suggest significant $\delta A_{LR}$, such as gauge-Higgs model\cite{Haba:2004qf} with tree level parity violation we will also analyze them.}


\vspace{1cm}

{\large \bf Acknowledegements}\\

\noindent
We thank 
 K. Hanagaki, T. Sato, K. Hikasa, K. Hagiwara, 
 C. S. Lim,  M. Wakamatsu, Y. Kuno, T. Yamanaka,   
 T. Onogi, and T. Kishimoto for useful and helpful discussions. 
S.T. is also grateful to H. Tanaka for helping the code implementation 
      of the numerical calculation.
We also thank M. Tanaka and K. Nagasaki for collaboration 
 in early stage of this work. 
This work is partially supported by Scientific Grant by Ministry of 
 Education and Science, Nos. 20540272, 22011005, 20244028, and 21244036.  



\end{document}